\documentclass{elsart}

\usepackage{harvard}

\usepackage{graphicx}

\usepackage{amssymb}

\def\url#1{{\ttfamily\def\/{/\discretionary{}{}{}}#1}}

\begin{document}

\begin{frontmatter}
\title{Multi-frequency observations as a key to source and environment
parameters of FRII objects}


\author[Kaiser]{C.R. Kaiser\thanksref{crk}}

\thanks[crk]{E-mail: ckaiser@mpa-garching.mpg.de}

\address[Kaiser]{Max-Planck-Institut f{\"u}r Astrophysik,
Karl-Schwarzschild-Str.1, 85740 Garching, Germany}

\begin{abstract}
Our knowledge of the environments of radio-loud AGN is still
sketchy. However, to understand the jet phenomenon it is important to
know about the properties of the surroundings in which jets are formed
and evolve. Here I present an analytical model of the radio surface
brightness distribution of the large scale structure of FRII-type
radio sources. The `virtual maps' resulting from this model can be
compared with observed maps to obtain estimates for a range of source
properties from the model. These properties include parameters
describing the gas density distribution of the source environment, the
energy transport rate of the jets and the orientation angle of the
source jet axis with respect to the line of sight. The model is tested
using radio maps of Cygnus A for which there are independent
measurements of some of these parameters available in the
literature. The model estimates agree well with these
observations. Varying the resolution of the radio maps used in the
comparison does not change the results significantly.
\end{abstract}

\end{frontmatter}


\section{Introduction}

The properties of AGN and their environments are of crucial importance
for many aspects of astrophysics. In the early universe the space
density of AGNs was orders of magnitude greater than it is today
\citeaffixed{dp90,hs90}{e.g.}. At the same time evidence is
accumulating that AGNs at high redshift, at least those which are
radio-loud, reside in the most massive and most evolved systems of the
respective epoch \citeaffixed{blr98}{e.g.}. It is therefore of great
importance for models of structure formation in the universe to know
the properties of the environments of radio sources. 

The jets in powerful radio galaxies and radio-loud quasars are
presumably powered by accretion discs around supermassive black
holes. This general picture is well accepted (however, see Kundt, this
volume) but the details of exactly how the jet material is accelerated
are still not very well understood. In this respect it is important to
know how much energy is transported along the jets because this places
constraints on the efficiency with which energy is channeled from the
accretion disc into the jet flow. 

The hot, gaseous environments of radio sources can be observed
directly at X-ray frequencies. The temperature of this gas and its
density can be derived from this emission. Unfortunately, the limited
sensitivity and resolution of X-ray telescopes confines this method to
objects at low to, at best, intermediate redshifts (see Worrall, this
volume). The uncertainties introduced by the additional X-ray emission
from the AGN itself are increased further by the possible scattering
of some of the X-rays emitted by the AGN off the material in the
environment of the source \cite{bcsf99}. Also, the large scale radio
structure embedded in the hot gas modifies the X-ray emission
properties of this material \cite{ka99}.

Optical emission from the surroundings of the large scale structure of
radio galaxies is detected in some objects on scales $\le 100$ kpc and
is often found to be aligned with the radio source at redshifts beyond
$z \sim 0.6$ \citeaffixed{pm93}{e.g.}. If the flux of ionising photons
from the AGN is known in these objects, then it is possible to
determine the density of the emitting material. However, the optical
light is emitted by the warm phase ($\sim 10^4$ K) rather than the hot
phase ($\sim 10^7$ K) of the gas in the surroundings of radio
sources. The inferred volume filling factors of the warm phase are
small and so the properties of the bulk of the gas, which is hot,
cannot be determined from the optical emission. Additionally, the
properties of the warm gas phase may also be changed by the expansion
of the large scale structure of the radio source \cite{ksr99}.

The radio emission of the large scale structure itself may also be
used in the determination of the properties of the environment. The
material external to the radio lobes is modifying the polarisation
properties of this radiation via Faraday rotation. The Rotation
Measure (RM) can be used to infer the density of the material creating
this Faraday screen. However, the RM depends not only on the gas
density but also on the strength of the magnetic field within this
material which is usually not very well constrained. The Faraday
screen is also known to be far from uniform. It is spatially
correlated with the optical emission in some sources and the RM may
therefore also probe predominantly the warm gas phase rather then the
hot phase \cite{cmb90}.

Constraining the jet power, i.e. the energy transport rate of the
jets, from radio observations usually involves some estimate of the
total energy content of the large scale structure and of the age of
the source in question. In most cases it is assumed that the
population of relativistic electrons is uniformly distributed over the
radio lobes and aging of the initial population is neglected in the
energy estimates. Spectral index maps, which are sometimes used for
the same sources to estimate their spectral ages, clearly demonstrate
that this is a poor assumption. Furthermore, in almost all sources the
orientation of the radio structure to the line of sight is unknown
which leads to errors in the estimation of the volume occupied by the
radio plasma. On the other hand multifrequency radio observations are
available for many radio sources of type FRII at all redshifts. Using
these maps to determine the properties of the radio sources and their
environments would provide a wealth of information without the need of
additional extensive observational programmes in many different
wavebands.

In the following I will therefore present a model which attempts to
determine the properties of the jets and those of the environment of
FRII sources using only radio maps at two observing frequencies. The
model predicts the radio surface brightness distribution of the large
scale structure of FRII-type objects based on 8 model
parameters. These parameters include the energy transport rate of the
jets, the gas density distribution of the source environment and the
angle of the jet axis with the line of sight. These `virtual radio
maps' are then compared to observations and the best fitting model
parameters are found. To determine the quality of these estimates I
use radio maps of Cygnus A for which independent measurements of some
of these quantities exist in the literature.

\section{The model}

\subsection{The dynamical evolution of FRII sources}

The model presented here is based on a model for the dynamics of
FRII-type radio sources developed in Kaiser \& Alexander (1997). The
basic features of this model can be summarised as follows. The usual
geometry of jets embedded in a cocoon which in turn is surrounded by a
bow shock as proposed by Scheuer (1974) is assumed. Parts of the
cocoon are identified as the observable radio lobes. The jets are
assumed to be in pressure equilibrium with the cocoon. Because of the
high sound speed in this region, the pressure within the cocoon should
be constant \cite{ka97} away from the region right next to the hot
spots which implies that the radius of the jets should remain constant
throughout the cocoon. The forward expansion of the cocoon is balanced
by the ram pressure of the surrounding gas which is pushed aside. The
model does not involve an assumption concerning the expansion of the
cocoon perpendicular to the jet axis. This implies that the model
results are independent of the exact geometrical shape of the
cocoon. The gas density of the external medium the source is expanding
into is modeled with a power law:

\begin{equation}
\rho =\rho _o \left( r / a_o \right)^{-\beta},
\end{equation}

\noindent where $r$ is the radial distance from the centre of the
distribution. Note, that the external density is therefore described
by two parameters, the exponent of the power law $\beta$ and the
combination $\rho _o a_o^{\beta}$ of the central density, $\rho_o$,
and the core radius, $a_o$. This model then predicts the expansion of
the cocoon to be self-similar which is supported by observations
\cite{lw84,lms89}.

\subsection{Radio emission of the cocoon}

The radio synchrotron emission of the cocoon is radiated by
relativistic electrons and may be also positrons gyrating about the
magnetic field lines within the cocoon. The spectral distribution of
the radiation emitted by a given electron depends on its energy. The
energy of a given electron is constantly changing because of losses
due to the (approximately) adiabatic expansion of the cocoon,
synchrotron radiation itself and inverse Compton scattering of the
cosmic microwave background photons. Of course the relativistic
electrons may gain energy as well if efficient reacceleration
processes are at work within the cocoon. For simplicity it is assumed
in the following that the relativistic particles are only accelerated
once at the termination shock of the jet.

All of the processes that may change the evolution of the energy
spectrum of the radiating relativistic particles are time
dependent. They will therefore have changed this spectrum to a larger
extent in parts of the cocoon which were injected by the jets at
earlier times. To keep track of these changes it is necessary to split
the cocoon into smaller volume elements characterised by their
injection time, $t_i$, into the cocoon. The evolution of the energy
spectrum of relativistic particles within each element is then
followed separately. This technique is described in greater detail in
Kaiser, Dennett-Thorpe \& Alexander (1997).

To proceed it is assumed that the energy spectrum of the relativistic
particles is initially given by a power law, $n \propto \gamma ^{-p}$,
where $\gamma$ is the relativistic Lorentz factor. The spectrum is
cut-off at high energies corresponding to $\gamma _{max}$. The model
then allows to calculate the synchrotron emissivity of the volume
elements for a given source age. Summing contributions of all the
elements then yields the total luminosity of the cocoon of the source.

\subsection{Spatial distribution of the radio emission}

\begin{figure}
\begin{center}
\includegraphics*[height=12cm, angle=-90]{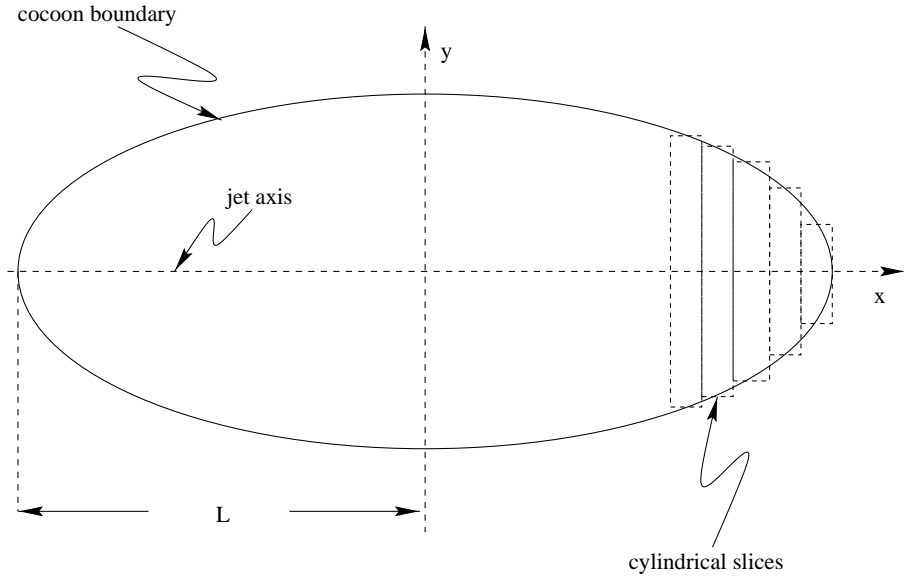}
\end{center}
\caption{The geometry of the cocoon.}
\label{fig:geo}
\end{figure}

The volume elements that are employed to follow the evolution of the
energy spectrum of the relativistic particles are `labeled' with their
injection time, $t_i$, but their spatial positions within the cocoon
are not specified. Therefore no assumption has been made so far
concerning the geometrical shape of the cocoon and it is possible to
choose a functional form for the surface delineating the cocoon
boundary which satisfies the shape of observed radio lobes. Assuming
rotational symmetry about the jet axis the parameterisation

\begin{equation}
y = L b_o \left( 1-l_x^2 \right) ^{b_1},
\end{equation}

in the plane containing the jet axis completely determines the cocoon
surface. Here $l_x$ is the dimensionless coordinate $l_x=x/L$ (see
Figure \ref{fig:geo}) with $L$ being the length of one jet. The
parameters $b_o$ and $b_1$ determine the aspect ratio of the cocoon
and the `bluntness' of its leading edge respectively.

In order to follow the evolution of the volume elements introduced
above not only in time but also in space, I now make the very
simplistic assumption that these elements can be identified with thin
cylindrical slices which are rotationally symmetric about the jet axis
(see Figure \ref{fig:geo}). This set-up was originally proposed by
Chy{\.z}y (1997) and has several implications. Firstly, the contents
of two neighboring slices are not allowed to mix. This assumption is
certainly too simplistic but because of the absence of pressure
gradients within the cocoon, which was assumed earlier, mixing of
thermal material and tangled magnetic field does not significantly
change the relevant conditions within the cocoon. Next, the possible
diffusion of relativistic particles does not significantly change
their energy distribution locally. This may hold, at least to some
extent, because the charged particles are tied to the magnetic field
lines which are assumed to be tangled on scales small compared to the
dimensions of the cocoon. Also the stochastic nature of the diffusion
process precludes the accumulation of relativistic particles of a
small energy range within a small volume and their depletion in the
rest of the cocoon. Finally, the individual slices must move as rigid
entities, i.e. the centre of a slice does not overtake its edge or
vice versa. Given these caveats, plus the assumption that there is no
significant reacceleration of relativistic particles in the cocoon, it
is probably best to view this model as describing the average
conditions within the cocoon at the position of a given slice.

The dynamical evolution underlying this emission model is self-similar
which implies for this model that all physical quantities behave as
power laws of age of the radio source. The dimensionless position of a
slice or volume element injected by the jet into the cocoon at time
$t_i$ is therefore set to $l_x=x/L \equiv ( t_i / t )^a$. The value of
$a$ is determined by the requirement that the sum of all volume
elements must be equal to the total volume of the cocoon.

Using this model it is now possible to create surface brightness plots
of FRII radio sources for a given set of source and environment
parameters by integrating the emissivity along the lines of sight
through the cocoon. To compare the model predictions with observations
the angle of the jet axis to the line of sight must also be
specified. In total there are 8 parameters in this model which
determine the predicted surface brightness distribution: $b_o$ and
$b_1$ which describe the unprojected geometrical shape of the cocoon,
$p$ and $\gamma _{max}$ determining the slope and cut-off of the
energy distribution of relativistic particles when they are injected
into the cocoon, the external density distribution is fixed by $\rho
_o a_o^{\beta}$ and $\beta$, the power of the jet $Q_o$ and the angle
of the jet to the line of sight, $\alpha _v$.

\section{Comparison with Cygnus A}

To test whether the model described above can correctly predict the
parameters describing the environment of an FRII source, we use Cygnus
A as a test case. For this source the properties of the gaseous
environment are known from X-ray observations \cite{cph94} and the
angle of the jets to the line of sight can be constrained from the jet
to counter-jet flux ratio \cite{hapr98}. 

Because of the model set-up it is possible to find the best fitting
parameters independently for the two radio lobes of Cygnus A. To find
the best fitting model parameters for one lobe we calculate the
surface brightness distribution for a given set of 8 parameters. The
age of the lobe is set by its observed length and the dynamical
model. The model prediction is then aligned with an observed lobe
using the core of the source. The `goodness of fit' is assessed by a
$\chi ^2$-like procedure which compares the observed map with the
model map pixel by pixel (the difference between model map and
observed map is squared and divided by the square of the 5$\sigma$
flux limit). The model parameters are then changed and the procedure
is repeated. The best fitting model parameters are found using a
8-dimensional downhill simplex method \cite{ptvf92} which minimises
the value obtained from the $\chi ^2$-like test. Note that in the
presence of discreet substructure in the lobes the flux measurements
in individual pixels are not independent of one another. This is
caused by the finite resolution of the beam of the telescope used in
the observations. The method of comparison used here is therefore not
a $\chi ^2$-test in the strict mathematical sense and must be viewed
as a maximum likelihood estimator. It is not possible to give an
absolute estimate of how well the observations are fitted by the
model.

In principle this fitting of model parameters can be done using one
observed map at a single frequency. However, because of degeneracies
in the model parameters involving the slope of the initial energy
spectrum of the relativistic particles, $p$, it is better to use two
maps at two different observing frequencies. The $\chi ^2$-values for
the two maps are then simply added together and the sum is minimised
by the fitting routine. 

\begin{figure}
\begin{center}
\includegraphics*[width=12cm, angle=-90]{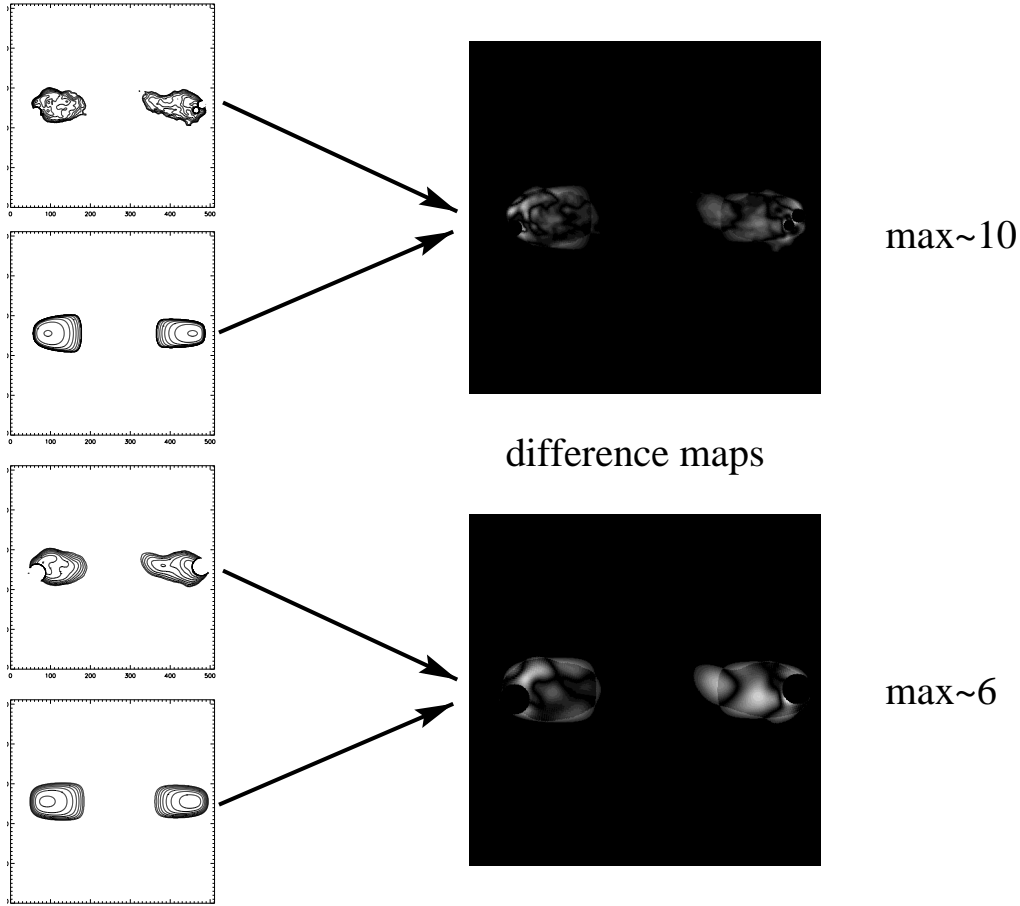}
\end{center}
\caption{Comparison of the model predictions with the observations of
Cygnus A. The contour plots show (from top to bottom) the high
resolution map at 1.7 GHz used to compare the model against, the model
prediction at high resolution, the convolved map at the same frequency
and the corresponding model prediction. The gray scale plots show the
difference between model and observations at high (top) and low
(bottom) resolution. The `max'-values given to the right indicate the
maximum deviation of the model in units of the 5$\sigma$ noise level
in the observed maps.}
\label{fig:res}
\end{figure}

For the comparison I obtained two maps of Cygnus A at 1.7 and 5 GHz
from the VLA archive. The angular resolution of both maps is
1.3$^{"}$. Because of the proximity of Cygnus A this corresponds to a
very high physical resolution which is usually not available for other
radio sources of such high power. To assess how the resolution of the
observations affect the model fit I performed two fits of each of the
two radio lobes, one at full resolution and one using the observed
maps convolved with a 5$^{"}$ beam. Any flux below the 5$\sigma$ limit
was removed from the maps. The model is not applicable to the hot spot
regions within the radio lobes and so the hot spot emission must be
removed from the observed maps. This was done by cutting out a
circular aperture centered on the peak of the surface brightness
distribution of each lobe. The radius of the aperture was set to the
distance from the aperture centre to the point where the 5$\sigma$
contour intersects the straight line running through the core of
Cygnus A and the centre of the aperture. In the high resolution case
an additional aperture removing the secondary hot spot in the western
lobe was used. The resulting observed maps for both resolutions are
shown on the left of Figure \ref{fig:res}. Only the lower frequency
maps are plotted but the results are very similar for the higher
frequency case.

Figure \ref{fig:res} also shows the best fitting model maps for both
radio lobes and a plot of the differences between model prediction and
observation. Note that for convenience both radio lobes are shown in
the same plot. However, the model parameters are somewhat different
for the two lobes (see Table \ref{tab:res}). The differences between
the model maps and the observed ones reach 50$\sigma$ in some
pixels. These large deviations occur mainly where discrete structure
is also seen in the observed maps. This underlines the claim that the
model describes the average conditions within the cocoon while it
fails in places where there are strong deviations from the mean. It is
therefore not surprising that the model fits the lower resolution maps
better than the maps with full resolution because the convolution with
a larger beam can be viewed to some extent as an averaging
process. Figure \ref{fig:res} also shows that the region where flux is
measured above the 5$\sigma$ limit in the western lobe extends further
along the jet axis then is predicted by the model. At the same time
the model prediction is `fatter' than the observed emission. The
western radio lobe of Cygnus A gives the impression of being squeezed
some way behind the hot spot. It is certainly narrower than its
eastern counterpart which is reflected in the model estimates of the
geometrical parameters $b_o$ and $b_1$ (see Table \ref{tab:res}). If
the lobe is squeezed, this may explain why radio plasma is detected
closer to the core. This material has simply a higher backflow
velocity than expected from the geometry of the lobe closer to the hot
spot. This illustrates that the model has problems in fitting
structures deviating from axisymmetry about the jet axis and/or
perturbed structures. Note however that the model estimates derived
from fitting the surface brightness distribution of the western lobe
are close to those obtained from the eastern lobe. The eastern lobe is
much more regular in appearance and the irregularities of the western
lobe therefore seem to be too weak to significantly alter the model
results.

Table \ref{tab:res} shows that there is little difference in the
derived model parameters between the high and the low resolution
fits. There is still enough information in the variation of the
surface brightness in the low resolution maps for the model to produce
good results. This is encouraging in the view of the difficulty of
obtaining high resolution radio maps of radio sources at high redshift
which are probably the most interesting candidates to apply the model
to in the future. Because of the large value of $p$, the model is, in
the case of Cygnus A, virtually independent of the high energy cut-off
of the energy distribution of the relativistic particles. $\gamma
_{max}$ is therefore not listed in Table \ref{tab:res}. 

Most importantly, the best fitting model parameters are close to the
ones derived from independent observations. The `variations' of the
model parameters given in Table \ref{tab:res} show how much a given
parameter can be varied before the $\chi ^2$-like estimator of the
goodness of fit changes by a factor $\sim 2$. The variation of the
viewing angle is large because the model only depends on the $\sin$ of
this angle and this does not change very much for the large viewing
angle of Cygnus A implied by the model. These variations can certainly
not be taken as proper error estimates for the predicted model
parameters. However, they indicate that the model presented here can
be used to produce reasonably firm estimates of the properties of
type-II radio sources and their environments.

\begin{table}
\begin{center}
\caption{Summary of model results for Cygnus A}
\begin{tabular}{llccccccc}
 & & b$_o$ & b$_{1}$ & p & $\beta$ & $\log (\rho _{o} a_{o} ^{\beta})^{\#}$ &
$\log (Q _{o}^{\ast})$ & $\alpha _{V}$\\
\hline
\vspace{0.5ex}
east & low res. & 0.24 & 0.17 & 2.3 & 1.42 & 6.5 & 39.1 & 89.8\\
     & high res.& 0.27 & 0.27 & 2.4 & 1.41 & 6.6 & 39.1 & 75.5\\[0.5ex]
west & low res. & 0.18 & 0.21 & 2.4 & 1.44 & 7.3 & 39.1 & 86.2\\
     & high res.& 0.17 & 0.17 & 2.4 & 1.44 & 7.3 & 39.1 & 76.4\\[0.5ex]
\multicolumn{2}{l}{`variation'} & 0.1 & 0.05 & 0.2 & 0.02 & 0.3 & 0.1 & 20.0 \\[0.5ex]
\multicolumn{2}{l}{measured} & & & & 1.45 & 7.0 & & 75.5\\
\hline
\end{tabular}
\end{center}
$^{\#}$ $\rho_o$ and $a_o$ are both measured in SI units.\\
$^{\ast}$ in Watts.
\label{tab:res}
\end{table}

\section*{Acknowledgments}

I would like to thank P. N. Best for patiently explaining the
intricacies of the reduction of radio interferometry data to a
theorist, i.e. me.

\end{document}